\documentclass[lettersize,journal]{IEEEtran}
\usepackage{amsmath,amsfonts}
\usepackage{algorithmic}
\usepackage{algorithm}
\usepackage{array}
\usepackage[caption=false,font=normalsize,labelfont=sf,textfont=sf]{subfig}
\usepackage{textcomp}
\usepackage{stfloats}
\usepackage{url}
\usepackage{verbatim}
\usepackage{graphicx}
\usepackage{cite}
\usepackage{bm}
\usepackage{amssymb}
\usepackage{multicol} 
\usepackage{caption}
\usepackage{booktabs}
\usepackage{multirow}

\begin{document}

\title{Model Predictive Contouring Control with Barrier and Lyapunov Functions for Stable Path-Following in UAV systems}

\author{Bryan S. Guevara, Viviana Moya, Luis F. Recalde, David Pozo-Espin, Daniel C. Gandolfo, Juan M. Toibero
\thanks{Bryan S. Guevara, Luis F. Recalde, Daniel C. Gandolfo and Juan M. Toibero are with the Instituto de Automática (INAUT), Universidad Nacional de San Juan-CONICET, San Juan J5400, Argentina (e-mail: bguevara@inaut.unsj.edu.ar; lrecalde@inaut.unsj.edu.ar; dgandolfo@inaut.unsj.edu.ar, mtoibero@inaut.unsj.edu.ar).}
\thanks{Viviana Moya is with the Facultad de Ingeniería, Universidad Internacional del Ecuador, Quito, 180101, Ecuador (e-mail: vivianamoya@uide.edu.ec).}
\thanks{David Pozo-Espin is with the Facultad de Ingeniería y Ciencias Aplicadas, Universidad de las Américas, Quito, 170513, Ecuador, and is the corresponding author (e-mail: david.pozo@udla.edu.ec).}}

\markboth{ IEEE Journal ,~Vol.~XX, No.~X, October~XXXX}%
{Shell \MakeLowercase{\textit{et al.}}: Model Predictive Contouring Control with Barrier and Lyapunov Functions for Stable Path-Following in UAV systems}

\IEEEpubid{0000--0000/00\$00.00~\copyright~2021 IEEE}

\maketitle

\begin{abstract}
In this study, we propose a novel method that integrates Nonlinear Model Predictive Contour Control (NMPCC) with an Exponentially Stabilizing Control Lyapunov Function (ES-CLF) and Exponential Higher-Order Control Barrier Functions to achieve stable path-following and obstacle avoidance in UAV systems. This framework enables unmanned aerial vehicles (UAVs) to safely navigate around both static and dynamic obstacles while strictly adhering to desired paths. The quaternion-based formulation ensures precise orientation and attitude control, while a robust optimization solver enforces the constraints imposed by the Control Lyapunov Function (CLF) and Control Barrier Functions (CBF), ensuring reliable real-time performance. The method was validated in a Model-in-the-Loop (MiL) environment, demonstrating effective path tracking and obstacle avoidance. The results highlight the framework’s ability to minimize both orthogonal and tangential errors, ensuring stability and safety in complex environments.

\end{abstract}


\section{Introduction}
\IEEEPARstart{T}{he} increasing complexity of modern aerial systems has driven substantial advancements in control methodologies capable of managing the non-linear dynamics and safety-critical constraints of Unmanned Aerial Vehicles (UAVs) \cite{Kumar2020}. UAVs are required to operate in dynamic and uncertain environments where they must not only follow predefined paths but also respond adaptively to the presence of obstacles and interactions with other vehicles \cite{Gurriet2020}. These scenarios place significant demands on control systems, which require simultaneous optimization of trajectory accuracy, safety, and real-time adaptability \cite{Gu2023}.

NMPCC has emerged as a promising control framework to address these challenges \cite{Romero2022,Romero2022_1, Lam2010}. Unlike traditional non-linear model predictive control (NMPC), which focuses on minimizing time-indexed tracking errors, NMPCC emphasizes minimizing contouring and lag errors relative to a desired path \cite{Xin2024}. This path-following optimization provides greater flexibility for UAVs to adapt their trajectories in response to environmental changes, such as the sudden appearance of obstacles, while still maintaining overall mission objectives \cite{Lam2013}. This adaptability makes NMPCC particularly well suited for complex tasks that require real-time responsiveness and high levels of autonomy \cite{Brito2019}.

The integration of NMPCC in UAV systems introduces additional challenges when operating in complex airspaces, requiring avoidance of both static and dynamic obstacles \cite{Krinner2024}. Ensuring safe, collision-free navigation under these conditions requires advanced control strategies that can dynamically adjust trajectories while maintaining the required stability and safety standards \cite{Wang2019}. Traditional control methods, though effective in simpler environments, often lack the robustness and flexibility to handle the complexity of multi-UAV systems in real-world applications \cite{Du2024,Niu2022,Huang2022}.

To address these limitations, we propose integrating NMPCC with CLF and CBF frameworks. CLFs offer a systematic numerical approach to ensuring system stability by enforcing the decrease of a Lyapunov function over time \cite{Wang2023, Ames2014, Ames2019, Garg2024}. On the other hand, CBFs enforce safety constraints, ensuring that UAVs operate within safe boundaries and avoid collisions \cite{Jankovic2018, Choi2021}.
The combination of these control frameworks within the NMPCC structure results in a comprehensive approach that addresses both performance and safety, enabling UAVs to navigate complex and dynamic environments autonomously.
\IEEEpubidadjcol
In addition to path-following and safety considerations, three-dimensional orientation is crucial for UAV control, particularly in dynamic environments where precise maneuvers are required \cite{Zheng2023}. To address this, we incorporate quaternion-based formulations into the control design. Quaternions offer several advantages over traditional Euler angles, such as avoiding gimbal lock and ensuring smooth, continuous rotation \cite{Huynh2009}. By leveraging the mathematical properties of quaternions, particularly their ability to map to the tangent space using the logarithmic map (Log) \cite{Tang2024,sola2021microlietheorystate, alcan2023trajectoryoptimizationmatrixlie}, we can effectively manage the rotational dynamics of UAVs, enhancing the robustness and accuracy of the overall control system. This quaternion-based approach is rooted in Lie theory and manifold principles, providing a mathematically rigorous foundation for handling the rotational behavior of UAVs.

Implementing this integrated framework, which combines NMPCC, CLFs, CBFs, and quaternion-based formulations, introduces considerable computational challenges, especially for real-time applications. To mitigate these challenges, we utilize ACADOS \cite{acados}, an open-source software package optimized for solving optimal control problems with high computational efficiency, alongside CasADi \cite{CasADi}, a symbolic framework for automatic differentiation and numerical optimization. These frameworks are particularly well-suited for UAV control tasks that demand rapid and precise decision-making, offering flexibility that enables seamless integration with the proposed control strategy. For the numerical integration of rotational dynamics, a fourth-order Runge-Kutta method is applied to ensure precise state updates.

In this study, we validate the proposed control strategy through extensive Model-in-the-Loop (MiL) simulations. These simulations demonstrate the practicality and robustness of the approach in complex environments. The results highlight the benefits of integrating NMPCC with advanced control techniques, offering a unified solution for safe and reliable UAV navigation in challenging operational settings.

In summary, the contributions of this paper are threefold:
\begin{itemize}
\item We present a novel integration of NMPCC with Control ES-CLF and higher-order CBF, providing a unified approach to dynamic obstacle avoidance in environments with multiple obstacles.
\item We demonstrate the effectiveness of quaternion-based formulations, using the Log operator to map to the tangent space according to Lie theory and manifolds, improving the robustness and precision of UAV attitude control by ensuring smooth transitions and accurate orientation representation.
\end{itemize}

\subsection{Outline}
This paper is structured as follows. Section II introduces the dynamic model of the UAV system, providing the foundation for the control strategy. Section III presents the CLF used to ensure stability in the proposed framework. Section IV focuses on the CBF, which enforce safety constraints in the system. Section V explores the NMPCC approach, detailing its application to path-following and obstacle avoidance. Section VI provides an in-depth analysis of the experiments and results, demonstrating the performance of the proposed control strategy. Finally, Section VII concludes the paper, summarizing the key findings and outlining potential directions for future research.

\section{Kinodynamic of UAV}

We base the kinodynamic model of the UAV described in \cite{Romero2022_}, where the state vector of the UAV is defined as $
\mathbf{x} = \begin{bmatrix} \mathbf{p} ~
\mathbf{v} ~
\mathbf{q} ~
\bm{\omega}
\end{bmatrix}^\intercal$, where \( \mathbf{p} = [x, y, z]^\intercal \) represents the UAV's position in the global frame, and \( \mathbf{v} = [v_x, v_y, v_z]^\intercal \) is the velocity vector also in the global frame. The orientation of the UAV is represented by the quaternion \( \mathbf{q} = \begin{bmatrix} q_w & \mathbf{q}_v^\intercal \end{bmatrix}^\intercal \), which defines defines the UAV's attitude. Finally, the angular velocity vector in the body frame is denoted by \( \bm{\omega} = [\omega_x, \omega_y, \omega_z]^\intercal \).

\subsection{Translational Dynamics}

The translational dynamics of the UAV is governed by the following equations:
\[
\dot{\mathbf{p}} = \mathbf{v},
\]
\[
\dot{\mathbf{v}} = \mathbf{g} + \frac{1}{m} \mathbf{R}\mathbf{F},
\]
where \( \mathbf{F} = \begin{bmatrix} 0 & 0 & F_z \end{bmatrix}^\intercal \) is the thrust force vector, with \( F_z \) representing the vertical thrust component. The rotation matrix \( \mathbf{R} \) transforms the vectors from the body frame to the global frame, while the gravitational acceleration vector \( \mathbf{g} = \begin{bmatrix} 0 & 0 & -g \end{bmatrix}^\intercal \) includes \( g \), the acceleration due to gravity. The mass of the UAV is denoted by \( m \).

\subsection{Rotational Dynamics}

The rotational dynamics of the UAV is described by the following equation:
\[
\dot{\bm{\omega}} = \mathbf{I}^{-1} \left( \bm{\tau} - \bm{\omega} \times (\mathbf{I} \cdot \bm{\omega}) \right),
\]
where \( \bm{\tau} = \begin{bmatrix}
    \tau_x & \tau_y & \tau_z
\end{bmatrix} \) is the moment vector applied to the UAV, \( \mathbf{I} \) is the inertia matrix, and \( \dot{\bm{\omega}} \) is the angular acceleration vector. The term \( \bm{\omega} \times (\mathbf{I} \cdot \bm{\omega}) \), known as the gyroscopic torque, accounts for the rotational effects due to angular velocity.

The evolution of the UAV’s attitude over time, which represents the instantaneous kinematics, is given by the quaternion derivative with respect to time. In this context, \( \otimes \) denotes quaternion multiplication:

\[
\dot{\mathbf{q}} = \frac{1}{2} \mathbf{q} \otimes \bm{\omega}.
\]

This can also be expressed as:

\begin{equation}
    \dot{\mathbf{q}}(t) = \frac{1}{2} 
    \begin{bmatrix}
        0 & -\bm{\omega}^\intercal (t)\\
        \bm{\omega}(t) & \begin{bmatrix}
            \bm{\omega}(t)
        \end{bmatrix}_\times
    \end{bmatrix} \mathbf{q}(t),
\end{equation}
or equivalently:

\begin{equation}
    \dot{\mathbf{q}}(t) = \frac{1}{2} 
    \mathbf{S}(\omega(t)) \mathbf{q}(t),
\end{equation}
where \( \begin{bmatrix} \bm{\omega}(t) \end{bmatrix}_\times \) is a skew-symmetric matrix.

\section{Control Lyapunov function}

Consider the UAV nonlinear dynamical system described by:
\begin{equation}\label{eq:sistema_general}
\dot{\bm{x}} = f(\bm{x}) + g(\bm{x})\bm{u},
\end{equation}
where \(\bm{x} \in \mathbb{R}^n\) is the state vector, \( f(\bm{x}): \mathbb{R}^n \rightarrow \mathbb{R}^n \) represents the uncontrolled dynamics, and \( g(\bm{x}): \mathbb{R}^n \rightarrow \mathbb{R}^{n \times m} \) is the control distribution matrix, defining how the control inputs \( \bm{u} \in \mathbb{R}^m \) affect the system.

A scalar function \(V: \mathbb{R}^n \rightarrow \mathbb{R}\) is called a \textit{Control Lyapunov Function} if it is continuously differentiable, positive definite, and satisfies \( V(\bm{x}_e) = 0 \), where \( \bm{x}_e \) represents the equilibrium point. This indicates that the system is at equilibrium when \( V(\bm{x}) = 0 \), and away from equilibrium when \( V(\bm{x}) > 0 \) for all \( \bm{x} \in \mathbb{R}^n \setminus \{\bm{x}_e\} \).

For \( V(\bm{x}) \) to be effective as a CLF, its time derivative along the system trajectories under the control law \( \bm{u} \) should be negative definite. Mathematically, this requirement is expressed as:
\begin{equation}
\dot{V}(\bm{x}) = \nabla V(\bm{x}) \cdot \dot{\bm{x}} < 0 \quad \text{for all } \bm{x} \in \mathbb{R}^n \setminus \{\bm{x}_e\}.
\end{equation}

This condition implies that the function \(V(\bm{x})\) decreases strictly along the trajectories of the system for an appropriate choice of the control input \(\bm{u}\), leading to the asymptotic stability of the system at the equilibrium point \( \bm{x}_e \).

An \textit{Exponentially Stabilizing Control Lyapunov Function} is a specialized form of a CLF that ensures exponential convergence of the state vector \(\bm{x}\) to the equilibrium point \(\bm{x}_e\). The function \(V(\bm{x})\) is classified as an ES-CLF if it satisfies the following conditions \cite{Ames2014}:

\begin{itemize}
    \item There exist positive constants \(c_1, c_2 > 0\) such that:
    \begin{equation}
    c_1 \|\bm{x} - \bm{x}_e\|^2 \leq V(\bm{x}) \leq c_2 \|\bm{x} - \bm{x}_e\|^2, \quad \text{for all} \, \bm{x} \in \mathbb{R}^n.
    \end{equation}

    \item There exists a positive constant \(c_3 > 0\) and a continuous control law \(\bm{u} = \bm{k}(\bm{x})\) that ensures the following condition holds:
    \begin{equation}
    \dot{V}(\bm{x}) \leq -c_3 V(\bm{x}), \quad \text{for all} \, \bm{x} \in \mathbb{R}^n, \, \bm{x} \neq \bm{x}_e.
    \end{equation}
\end{itemize}

Integrating this inequality with respect to time yields:
\begin{equation}
V(\bm{x}(t)) \leq V(\bm{x}(0)) e^{-c_3 t}.
\end{equation}

Given the quadratic bounds on \(V(\bm{x})\), we can further deduce:
\begin{equation}
\resizebox{\linewidth}{!}{$
c_1 \|\bm{x}(t) - \bm{x}_e\|^2 \leq V(\bm{x}(t)) \leq V(\bm{x}(0)) e^{-c_3 t} \leq c_2 \|\bm{x}(0) - \bm{x}_e\|^2 e^{-c_3 t}.
$}
\end{equation}

Taking the square root of both sides and rearranging terms, we obtain:
\begin{equation}
\|\bm{x}(t) - \bm{x}_e\| \leq \sqrt{\frac{c_2}{c_1}} \|\bm{x}(0) - \bm{x}_e\| e^{-\frac{c_3}{2} t}.
\end{equation}

This condition implies that \(V(\bm{x})\) decays exponentially over time, guaranteeing that the state vector \(\bm{x}\) converges to the equilibrium point \(\bm{x}_e\) exponentially, with a rate of convergence determined by the constant \(c_3\).

The use of an ES-CLF as a Lyapunov function in control design carries significant theoretical stability guarantees \cite{Ames2013, Minniti2022}. This exponential stability implies that the system's state vector \(\bm{x}(t)\) converges to the equilibrium point \( \bm{x}_e \) at an exponential rate, which is a stronger form of stability compared to mere asymptotic stability.

\section{Control Barrier Functions }

A \textit{Control Barrier Function} is defined for a safe set \(\mathcal{C} \subset \mathbb{R}^n\) as:
\begin{equation}
\mathcal{C} = \{\bm{x} \in \mathbb{R}^n \mid h(\bm{x}) \geq 0\},
\end{equation}
where \(h: \mathbb{R}^n \rightarrow \mathbb{R}\) is a twice continuously differentiable scalar function.

To ensure that the system state remains within the safe set \(\mathcal{C}\), we consider the successive Lie derivatives of \(h(\bm{x})\) along the system dynamics. The Lie derivative of \(h(\bm{x})\) with respect to \(f(\bm{x})\) is defined as:
\begin{equation}
L_f h(\bm{x}) = \nabla h(\bm{x}) \cdot f(\bm{x}),
\end{equation}
and the Lie derivative of \(h(\bm{x})\) with respect to \(g(\bm{x})\) is:
\begin{equation}
L_g h(\bm{x}) = \nabla h(\bm{x}) \cdot g(\bm{x}).
\end{equation}

Here, \(\nabla h(\bm{x})\) is the gradient of the function \(h(\bm{x})\) with respect to the state variables \(\bm{x}\), and this gradient is multiplied by \(f(\bm{x})\) and \(g(\bm{x})\), respectively.

The Lie derivative of order \(k\) of the function \(h(\bm{x})\) along the system dynamics is:
\begin{align}
L_f^k h(\bm{x}) &= \frac{d^k h(\bm{x})}{d t^k}  \\
                &= \frac{\partial}{\partial \bm{x}} \left( L_f^{k-1} h(\bm{x}) \right) f(\bm{x}).
\end{align}

Additionally, the cross Lie derivative of order \(k-1\) involving the function \(g(\bm{x})\) is expressed as:
\begin{equation}
L_g L_f^{k-1} h(\bm{x}) = \nabla \left( L_f^{k-1} h(\bm{x}) \right) \cdot g(\bm{x}),
\end{equation}
where \(L_f^{k-1} h(\bm{x})\) is the \((k-1)\)-th order Lie derivative of \(h(\bm{x})\) with respect to \(f(\bm{x})\).

To handle more complex safety constraints, we introduce higher-order Control Barrier Functions, which are governed by the following condition:
\begin{equation}
L_f^k h(\bm{x}) + L_g L_f^{k-1} h(\bm{x}) \bm{u} + \dots + L_g h(\bm{x}) \bm{u} \geq -\alpha_k(h(\bm{x})),
\end{equation}
where \(\alpha_k: \mathbb{R} \rightarrow \mathbb{R}\) is a class \(\mathcal{K}\) function, meaning it is continuous, strictly increasing, and \(\alpha_k(0) = 0\).

The complete expression for \(k\) derivatives can be expanded as:
\begin{equation}
\sum_{i=0}^{k} \binom{k}{i} L_f^{k-i} L_g^i h(\bm{x}) \bm{u} \geq -\alpha_k(h(\bm{x})),
\end{equation}
where \(\binom{k}{i}\) is the binomial coefficient. This condition ensures that, under the action of the control input \(\bm{u}\), the system will remain within the safe set \(\mathcal{C}\) defined by the function \(h(\bm{x})\).

For \(h(\bm{x})\) to be an \textit{Exponential Control Barrier Functions}, there must exist a gain vector \(\mathbf{K}_\alpha = [K_{\alpha,1}, K_{\alpha,2}, \dots, K_{\alpha,r}] \in \mathbb{R}^r\) such that for the system described by (\ref{eq:sistema_general}), the following condition holds:
\begin{equation}
\sup_{\bm{u} \in U} \left[ L_f^r h(\bm{x}) + L_g L_f^{r-1} h(\bm{x}) \bm{u} \right] \geq -\mathbf{K}_\alpha \eta_b(\bm{x}),
\end{equation}
where the vector \(\eta_b(\bm{x})\) groups the derivatives of \(h(\bm{x})\) and its Lie derivatives up to order \(r-1\). The vector \(\eta_b(\bm{x})\) is constructed as:
\begin{equation}
\eta_b(\bm{x}) = \begin{pmatrix} 
h(\bm{x}) \\ 
L_f h(\bm{x}) \\ 
L_f^2 h(\bm{x}) \\ 
\vdots \\ 
L_f^{r-1} h(\bm{x}) 
\end{pmatrix}.
\end{equation}

To compute the second-order Lie derivative of the barrier function, we apply the Lie derivative to \(L_f h(\bm{x})\) as follows:
\begin{equation}
L_f^2 h(\bm{x}) = \nabla (L_f h(\bm{x})) \cdot f(\bm{x}),
\end{equation}
\begin{equation}
L_g L_f h(\bm{x}) = \nabla (L_f h(\bm{x})) \cdot g(\bm{x}).
\end{equation}

This approach allows for the calculation of barrier functions up to the second order, which is essential for implementing constraints of the form discussed:
\begin{equation}
\ddot{h}(\bm{x},\bm{u} ) = L_f^2 h(\bm{x}) + L_g L_f h(\bm{x}) \bm{u}.
\end{equation}

\section{Model Predictive Contouring Control}

Unlike traditional NMPC, the formulation of \textit{Nonlinear Model Predictive Contouring Control} is used to minimize trajectory tracking errors while maximizing the speed along a given path over a finite prediction horizon \( l \in [t, t+N] \) \cite{Lam2013}. This approach is particularly useful in three-dimensional (3D) environments, where the objective is to reduce both the contour error (orthogonal to the trajectory) and the lag error (parallel to the trajectory), while ensuring efficient progress along the path.

\begin{figure}[t]
\centering
\includegraphics[clip, trim=12cm 6cm 14.5cm 6cm, width=0.9\linewidth]{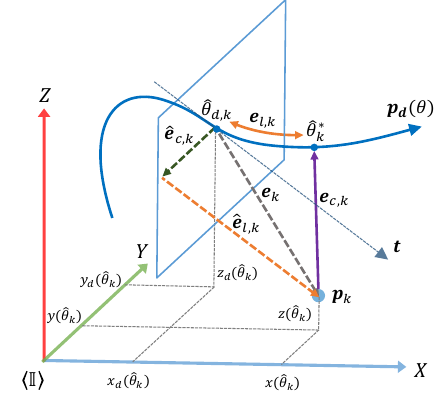}
\caption{ The position error is projected onto orthogonal vectors, resulting in approximations of the contour and lag errors.}
\label{Fig:Diagrama_MPCC}
\end{figure}

The desired trajectory is defined as a smooth curve parametrized by the arc length $\theta$, also referred to as \textit{progress} along the path. The position at any point on the trajectory is given by $\mathbf{p}_d(\theta) = \begin{bmatrix} x_d(\theta) ~ y_d(\theta) ~ z_d(\theta) \end{bmatrix}^\intercal$, and its derivative with respect to \( \theta \), which corresponds to the tangent vector, is:
\[
\mathbf{t}(\theta) = \mathbf{p}'_d(\theta) = \frac{d}{d\theta} \mathbf{p}_d(\theta).
\]

Since the trajectory is parametrized by arc length, the norm of the tangent vector is always one:
\[
\|\mathbf{t}(\theta)\| = 1,
\]

At any time \( t_k \), the point \( \theta_k^* \) represents the location on the desired path closest to the current position \( \mathbf{p}(t_k) \). However, since \( \theta_k^* \) is typically unknown during optimization, the system approximates it with \( \hat{\theta}_k \), which serves as the best estimate of \( \theta_k^* \). The total position error is defined as the difference between the current system position \( \mathbf{p}(t_k) \) and the desired position \( \mathbf{p}_d(\hat{\theta}_k) \) on the trajectory:
\[
\mathbf{e}(\hat{\theta}_k) = \mathbf{p}(t_k) - \mathbf{p}_d(\hat{\theta}_k).
\]

Although the system position is time-parametrized and the desired position is progress-parametrized, the comparison is valid because \( \hat{\theta}_k \) corresponds to the position on the trajectory at \( t_k \). As shown in Fig. \ref{Fig:Diagrama_MPCC}, the total error can be decomposed into two orthogonal components: the \textit{lag error}, \( \mathbf{e}_l(\hat{\theta}_k) \), and the \textit{contour error}, \( \mathbf{e}_c(\hat{\theta}_k) \).

The lag error is defined as the difference in arc length between the current position \( \theta_k \) and the real minimizer \( \theta_k^* \):
\[
e_l(\theta_k^*) = \theta_k - \theta_k^*.
\]

It is often approximated by projecting the total position error \( \mathbf{e}(\hat{\theta}_k) \) onto the tangent vector \( \mathbf{t}(\hat{\theta}_k) \):
\[
\mathbf{e}_l(\hat{\theta}_k) = \left(\mathbf{e}(\hat{\theta}_k) \cdot \mathbf{t}(\hat{\theta}_k)\right) \mathbf{t}(\hat{\theta}_k),
\]
which results in a vector that represents the component of the total error along the direction of motion. Here, \( \mathbf{t}(\hat{\theta}_k) = \mathbf{p}'_d(\hat{\theta}_k) \).

On the other hand, the contour error quantifies the deviation perpendicular to the trajectory. It is calculated by subtracting the lag error from the total position error. Alternatively, the contour error can be computed by projecting the position error onto the normal to the tangent vector. This projection is performed using the matrix \( P_{ec} \), which is defined as:
\[
P_{ec} = I - \mathbf{t}(\hat{\theta}_k) \mathbf{t}^\intercal(\hat{\theta}_k),
\]
where \( I \) is the identity matrix. The contour error is then given by:
\[
\mathbf{e}_c(\hat{\theta}_k) = P_{ec}~\mathbf{e}(\hat{\theta}_k).
\]

The contour error can also be understood as the solution to the following minimization problem:
\[
e_c(\hat{\theta}_k) = \min_{\theta} \| \mathbf{p}_k - \mathbf{p}_d(\theta) \|.
\]

If the projection of the error onto the tangent direction \( \mathbf{e}_l(\hat{\theta}_k) \) is zero, the approximations \( \hat{\theta}_k \), \( \mathbf{e}_c(\hat{\theta}_k) \), and \( \mathbf{e}_l(\hat{\theta}_k) \) coincide with their true values, ensuring accurate tracking.

Finally, the \textit{velocity of progress} \( v_{\hat{\theta},k} \) quantifies the rate at which the system moves along the trajectory. It is calculated by projecting the system’s velocity \( \mathbf{v}_k \) onto the tangent vector \( \mathbf{t}(\hat{\theta}_k) \):
\[
v_{\hat{\theta},k} = \mathbf{v}_k \cdot \mathbf{t}(\hat{\theta}_k),
\]
which gives the instantaneous speed at which the system progresses along the tangent path.

\subsection{Optimization Problem for NMPCC}
In addition to minimizing deviations from the desired path and control effort, the NMPCC also considers maintaining the system's attitude orientation aligned with the direction of motion along the path.

This approach leverages the logarithm map \( \text{Log}(\cdot) \), which maps the quaternion error from the unit quaternion space \( \mathbb{H} \) onto the tangent space \( \mathbb{R}^3 \). This map is defined as:\begin{equation}
\text{Log}(\Tilde{\mathbf{q}}) = 2 \Tilde{\mathbf{q}}_v \cdot \frac{\text{atan2}(\left\| \Tilde{\mathbf{q}}_v \right\|, \Tilde{q}_w)}{\left\| \Tilde{\mathbf{q}}_v \right\|},
\end{equation}
where \( \Tilde{q}_w \in \mathbb{R} \) is the scalar component and \( \Tilde{\mathbf{q}}_v \in \mathbb{H}_p\) is the vector part of the quaternion error. This transformation is particularly useful for optimizing orientation errors in Euclidean space. The orientation error between a desired quaternion \( \mathbf{q}_d \) and the actual quaternion \( \mathbf{q} \) is computed as:\begin{equation}
\Tilde{\mathbf{q}} =  {\mathbf{q}}_d \otimes {\mathbf{q}}^{-1},
\end{equation}
where \( \otimes \) denotes quaternion multiplication. The inverse of a quaternion can be calculated using its conjugate, denoted as \( \bar{\mathbf{q}} \), along with its norm. Specifically, the inverse is given by:\begin{equation}
\mathbf{q}^{-1} = \frac{\bar{\mathbf{q}}}{||\mathbf{q}||^2},
\end{equation}

The quaternion manifold \( \mathcal{S}^3 \), representing unit quaternions, is a double cover of the special orthogonal group \( SO(3) \), meaning that both \( \mathbf{q} \) and \( -\mathbf{q} \) describe the same rotation \cite{Jackson2021}. To avoid ambiguity arising from this property, it is essential to ensure that the real part \( q_w \) is positive; if \( q_w \) is negative, the quaternion should be negated before proceeding with any computations.
The optimization problem is then formulated as:
\begin{equation}
\begin{split}
\min_{\mathbf{u_k}, \Tilde{\mathbf{q}}_k} & \quad \sum_{k=0}^{N}  \| \mathbf{e}_c(\hat{\theta}_k) \|^2_{\mathbf{Q}_c}  
+ \| \mathbf{e}_l(\hat{\theta}_k) \|^2_{\mathbf{Q}_l} - \mu v_{\hat{\theta}, k}^2 + \\
&  \quad \left\| \text{Log} (\Tilde{\mathbf{q}}_k) \right\|^2_{\mathbf{Q}_q} 
+ \frac{1}{2} \left\| \mathbf{u}_k  \right\|^2_\mathbf{R} + \rho \zeta^2\\
\text{subject to: } & \quad \bm{x}_{k+1} = f(\bm{x}_k) + g(\bm{x}_k)\bm{u}_k,\\
& \quad \bm{x}_0 = \bm{x}(0), \\
& \quad \mathbf{u}_k \in \mathbb{U},  \\
& \quad \bm{x}_k \in \mathbb{X}, \\
& \quad 0 \leq v_{\theta_k} \leq v_{\theta_\text{max}}, \\
& \quad \dot{V}(\bm{x}_k) \leq -\gamma {V}(\bm{x}_k) + \zeta, \\
& \quad \ddot{h}(\bm{x}_k, \bm\mu_k) \geq -\mathbf{K}_\alpha \eta_b(\bm{x}_k).
\end{split}
\end{equation}

The gains \( \mathbf{Q}_c \) and \( \mathbf{Q}_l \) represent the weights for the contour and lag errors, respectively. \( \mu \) adjusts the trade-off for higher progress speed \( v_{\hat{\theta}} \) to move as quickly as possible along the path, while \( \mathbf{Q_q} \) is the weight matrix for the quaternion error. Finally, \( \mathbf{R} \) penalizes the control effort;

In the context of NMPCC, a Lyapunov candidate function is introduced to ensure system stability while minimizing the error along the path. The Lyapunov candidate function \( V \) considers both the contour error \( \mathbf{e}_c(\hat{\theta}_k) \) and the lag error \( \mathbf{e}_l(\hat{\theta}_k) \), and is defined as:
\[
V = \frac{1}{2} \mathbf{e}_c^\top \mathbf{W}_c \mathbf{e}_c + \frac{1}{2} \mathbf{e}_l^\top \mathbf{W}_l \mathbf{e}_l,
\]
where \( \mathbf{W}_c \) and \( \mathbf{W}_l \) are positive definite weighting matrices for the contour and lag errors, respectively.

The time derivative of \( V \) is then given by:

\[
\dot{V} = \frac{\partial V}{\partial \mathbf{p}} \mathbf{v},
\]
where the gradient of this function \( \frac{\partial V}{\partial \mathbf{p}} \) is computed with respect to the position vector \( \mathbf{p} \), and \( \mathbf{v} \) is the velocity vector, previously defined.

To enforce the Lyapunov condition a slack variable \( \zeta \geq 0 \) is introduced:
\[
\dot{V} \leq -c_3 V + \zeta.
\]

The slack variable \( \zeta \) introduces flexibility when constraints are too strict, allowing the system to remain feasible. To limit its impact, \( \zeta \) is penalized in the cost function with the term \( \rho \zeta^2 \), where \( \rho > 0 \) controls the degree of penalization. This ensures feasibility in the Optimal Control Problem (OCP) while preserving stability. In real-time applications, the use of a slack variable becomes crucial for handling uncertainties and ensuring the optimization process remains feasible, especially under tight computational constraints.

The barrier function is designed to ensure that the distance between the UAV and the obstacle remains greater than the combined size of both, plus a safety margin. This function is expressed as:
\[
h = \|\mathbf{p} - \mathbf{p}_{obs}\| - \left( r_{uav} + r_{obs} + \text{margin} \right),
\]
where \( \mathbf{p}_{obs} = [x_{obs}, y_{obs}, z_{obs}]^T \) represents the position of the obstacle. The term \( r_{uav} \) corresponds to the UAV's radius, \( r_{obs} \) to the obstacle's radius, and the constant \text{margin} ensures a minimum additional distance to prevent collisions.

\subsection{Arc-Length Parametrization via the Bisection Method}

To parametrize the desired trajectory \( \mathbf{p}_d(t) \) by arc length, we use the \textit{bisection method} to determine the corresponding time \( t_k \) for each arc-length value \( \theta_k \). Given a continuous trajectory \( \mathbf{p}_d(t) \), the arc length from the initial point \( t_0 \) to any point \( t_k \) is given by the following equation:
\begin{equation}\label{eq:bisection}
\theta_k = L(t_k) = \int_{t_0}^{t_k} \left\| \frac{d}{dt} \mathbf{p}_d(t) \right\| \, dt,    
\end{equation}

Since there is no explicit expression for \( t_k \) as a function of \( \theta_k \), the bisection method is employed to numerically find \( t_k \) such that Eq. \ref{eq:bisection} is satisfied for the given \( \theta_k \). To do so, we initialize the bisection algorithm with an interval \( [t_{\text{low}}, t_{\text{high}}] \), where \( t_{\text{low}} \) and \( t_{\text{high}} \) are guesses that are expected to contain \( t_k \). Typically, \( t_{\text{low}} \) can be initialized as \( t_0 \), the starting point of the trajectory, and \( t_{\text{high}} \) is chosen based on an estimate of where the desired arc length \( \theta_k \) is reached.

The bisection method, as outlined in Algorithm \ref{alg:time_calculation}, iteratively refines the interval \( [t_{\text{low}}, t_{\text{high}}] \) to find the time \( t_k \) that corresponds to the desired arc length \( \theta_k \). This process continues until the difference between the computed arc length \( L(t_{\text{mid}}) \) and \( \theta_k \) falls below the specified tolerance \( \epsilon \).

\begin{algorithm}[H]
\caption{Time Calculation for \(\theta_k\) Based on Arc Length}
\begin{algorithmic}
\STATE \textbf{Input}: Desired arc length \( \theta_k \), initial interval \( [t_{\text{low}}, t_{\text{high}}] \), tolerance \( \epsilon \).
\WHILE{\( |L(t_{\text{mid}}) - \theta_k| \geq \epsilon \)}
    \STATE Compute the midpoint \( t_{\text{mid}} = \frac{t_{\text{low}} + t_{\text{high}}}{2} \).
    \STATE Calculate the cumulative arc length \( L(t_{\text{mid}}) = \int_{t_0}^{t_{\text{mid}}} \|\mathbf{p}'_d(t)\| \, dt \).
    \IF{\( L(t_{\text{mid}}) < \theta_k \)}
        \STATE Update \( t_{\text{low}} = t_{\text{mid}} \).
    \ELSE
        \STATE Update \( t_{\text{high}} = t_{\text{mid}} \).
    \ENDIF
\ENDWHILE
\STATE \textbf{Output}: \( t_k = t_{\text{mid}} \) such that \( L(t_k) \approx \theta_k \).
\end{algorithmic}
\label{alg:time_calculation}
\end{algorithm}

 Once \( t_k \) is found, the corresponding position \( \mathbf{p}_d(\theta_k) \) is obtained as:
\[
\mathbf{p}_d(\theta_k) = \mathbf{p}_d(t_k) = \begin{bmatrix} x_d(t_k) & y_d(t_k) & z_d(t_k) \end{bmatrix}^\intercal.
\]

\section{Experiments and Results}
This section presents the results of applying the proposed NMPCC framework to a UAV system. The framework was tested in simulations where the UAV moved at high speeds in the presence of both static virtual obstacles and a mobile obstacle, in order to evaluate its path-following and obstacle-avoidance capabilities.



\subsection{Experimental Setup}
We conducted the experiments using a PC station with an AMD Ryzen 7 3700x and 16GB RAM, utilizing CasADi with ACADOS for optimal control.
The controller was executed at 30 Hz with a prediction horizon of 30 steps (1 second). The UAV followed a figure-eight path, parametrized by arc length, with a maximum progress velocity $v_{\theta_\text{max}}$ of 6 m/s. The UAV, with a radius of 0.15 m corresponding to a 6-inch frame \cite{Foehn2022}, included a safety margin of 0.1 m. The contour and lag errors were weighted by \( \mathbf{Q}_c = 3I_{3\times3} \) and \( \mathbf{Q}_l = I_{3\times3} \), respectively. The progress velocity was regulated by a gain \( \mu = 0.1 \), the control effort was penalized with \( \mathbf{R} = \text{diag}[0.02, 200, 200, 200] \) and the attitude error was weighted by \( \mathbf{Q}_q = I_{3\times3} \). The Lyapunov constraint, \( \mathbf{W}_c = \mathbf{W}_l = I_{3\times3} \) and \( \rho = 100 \) were used. A gain of \( \gamma = 0.9 \) and \( \mathbf{K}_\alpha = [20, 8]^\intercal \) for the static obstacle and \( \mathbf{K}_\alpha = [20, 15]^\intercal \) for the mobile obstacle were applied. The following reference path was considered: \( x(t) = 4 \sin(0.04 \cdot t) + 1 \), \( y(t) = 4 \sin(0.08 \cdot t) \), and \( z(t) = 2 \sin(0.08 \cdot t) + 6 \). The quadrotor was required to complete the experimental path within a fixed duration of 30 seconds.


\subsection{Simulated Performance }

\begin{figure}[H]
    \centering
    \includegraphics[clip, trim=0cm 0cm 0cm 0cm,width=\linewidth]{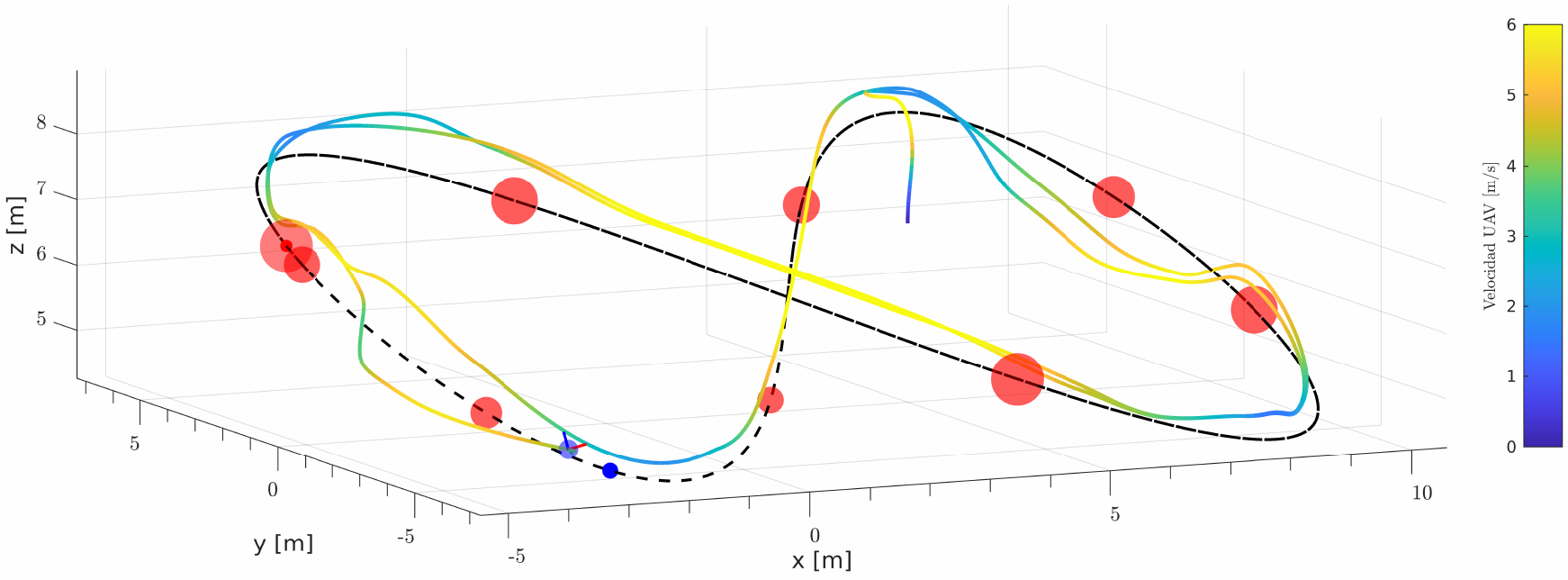}
    \caption{Flight path of UAV with MPCC in the presence of static obstacles.}
    \label{fig:static_obstacles}
\end{figure}

The MPCC framework was tested in simulations involving both static and mobile virtual obstacles to assess its path-following and obstacle-avoidance capabilities. During the test, a mobile obstacle moved along the same path but in the opposite direction, increasing the relative velocity between the UAV and the obstacle due to their opposing movements. Fig. \ref{fig:static_obstacles} illustrates the UAV's trajectory as it navigates around both static and mobile obstacles, demonstrating the MPCC's effectiveness in minimizing contour and lag errors while ensuring consistent progress along the desired path.

\begin{figure}[H]
    \centering
    \includegraphics[clip, trim=0cm 0cm 0cm 0cm,width=\linewidth]{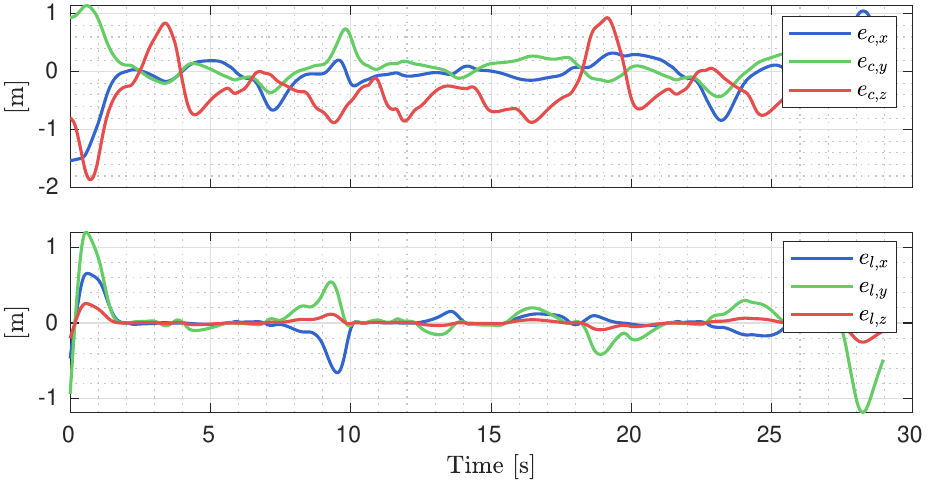}
    \caption{Contour and lag errors during path-following with obstacle avoidance.}
    \label{fig:Errors_static}
\end{figure}

The contour and lag errors recorded during the experiment are summarized in Fig. \ref{fig:Errors_static}, demonstrating the effectiveness of MPCC in maintaining low error values, with instantaneous contour errors remaining below \( \|\mathbf{e}_c\| < 1.97 \) m and instantaneous lag errors below \( \|\mathbf{e}_l\| < 1.87 \) m, when obstacles are present. It is observed that during obstacle avoidance maneuvers, the contour error is the most affected, as the UAV adjusts its path to navigate around the obstacle.

\begin{figure}[H]
    \centering
    \includegraphics[clip, trim=0cm 0cm 0cm 0cm,width=\linewidth]{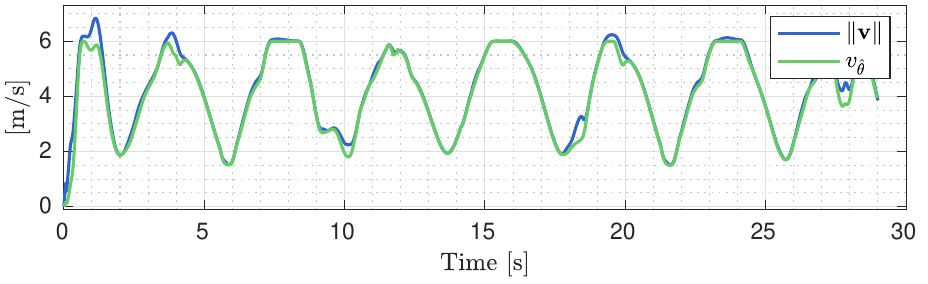}
    \caption{The system's instantaneous velocity and progress velocity.}
    \label{fig:velocity_dual}
\end{figure}

In Fig. \ref{fig:velocity_dual}, the progress velocities and the norm of the velocity are shown. While both curves appear similar due to the maintained direction, it's important to note that the progress velocity is the projection of the system’s velocity onto the reference’s tangent vector. Additionally, the progress velocity adheres to the constraints set in the OCP, ensuring controlled advancement along the trajectory.

\begin{figure}[H]
    \centering
    \includegraphics[width=\linewidth]{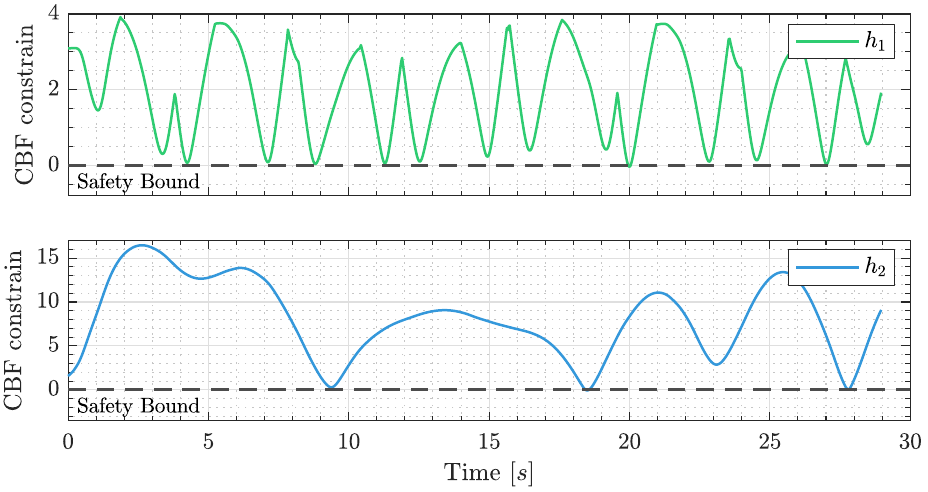}
    \caption{CBF safety constraints.}
    \label{fig:CBF_static}
\end{figure}

In Fig. \ref{fig:CBF_static}, the combined ECBF values ($h_1$ and $h_2$) are displayed, demonstrating their effectiveness in enforcing safety constraints along the UAV's trajectory, particularly around static and moving obstacles.

 By ensuring that \( h(x) > 0 \) throughout the experiment, the ECBF guarantees that the UAV avoids collisions and maintains safe operational limits throughout the flight. The minimum safety distance of between the obstacles was consistently maintained, with the CBF values remaining positive, thereby confirming that the safety measures were effectively enforced.

\begin{figure}[H]
    \centering
    \includegraphics[width=\linewidth]{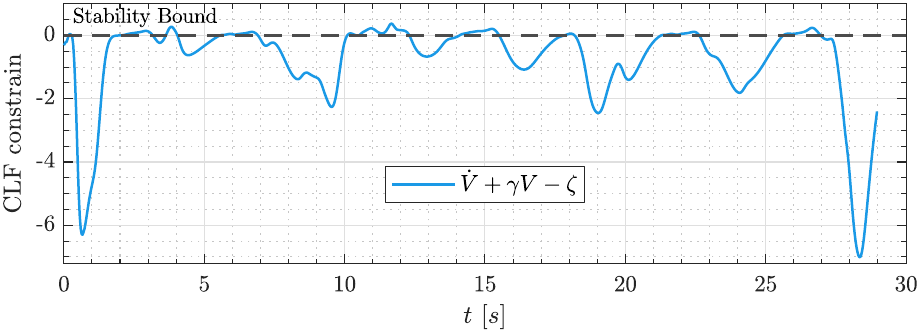}
    \caption{CLF stability constraints.}
    \label{fig:CLF_static}
\end{figure}

In Fig. \ref{fig:CLF_static}, the ES-CLF stability constraint ensures that the system remains stable. Temporary increases in this constraint are observed when the UAV performs corrective maneuvers, particularly during obstacle avoidance. However, the inclusion of a slack variable allows for minimal violations when necessary, ensuring that the optimizer remains feasible. This ensures that the system remains operational, and despite these temporary deviations, the stability metric trends toward convergence as the UAV returns to its desired path, highlighting the system's robustness and ability to regain stability after corrective actions.

\begin{figure}[H]
    \centering
    \includegraphics[width=\linewidth]{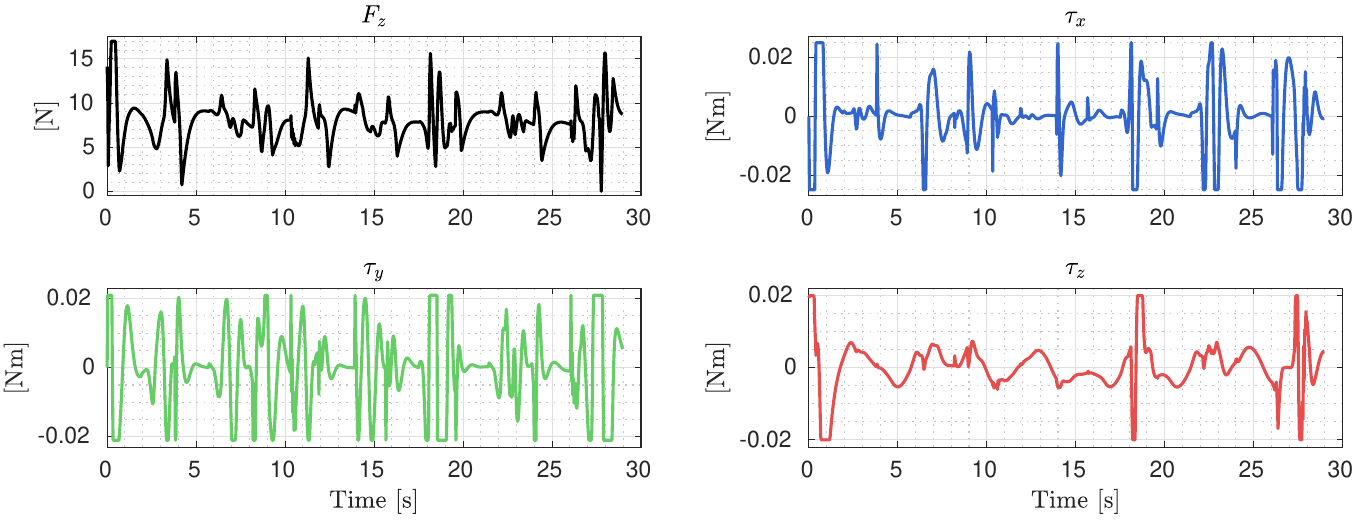}
    \caption{Control actions within OCP limits.}
    \label{fig:control_actions}
\end{figure}

In Fig.  \ref{fig:control_actions}, the control actions are shown, remaining within the limits set by the OCP. Notably, in the presence of obstacles, the control actions adjust effectively, ensuring obstacle avoidance without violating the imposed constraints. This illustrates the system's ability to handle complex scenarios while maintaining control stability.

\section{Conclusion}
The NMPCC framework provides an effective solution for tracking desired trajectories by minimizing both orthogonal and tangential errors while maximizing progress along the path. By integrating the trajectory geometry with the system dynamics, NMPCC enables precise and robust path tracking even in the presence of obstacles at high speeds.

The results from the ECLF and ECBF analyses confirm that the proposed NMPCC framework ensures both stability and safety during UAV path-following tasks. Future work will focus on extending this framework to multi-UAV systems and validating its performance in real-world flight tests.

Additionally, the NMPCC framework incorporates an attitude control strategy that uses the Log map to project quaternions into \(\mathbb{R}^3\). This enables the direct minimization of attitude errors in the cost function, allowing the integration of Lie algebra with standard Euclidean operations. As a result, the framework ensures precise alignment of the UAV’s orientation with the desired trajectory, permitting complex maneuvers to be handled effectively in geometric space.

\section*{Acknowledgments}
The authors are grateful for the support of INAUT - UNSJ, which provided them with the facilities and workspace necessary to carry out this research.

\bibliography{biblio.bib}
\bibliographystyle{IEEEtran}

\vspace{-1em}

\null \vfill

\end{document}